\documentclass[12pt]{iopart}
\usepackage[latin1]{inputenc}
\usepackage{graphicx,psfrag}

\newcommand{\average}[1]{\left<{#1}\right>}

\begin{document}
\title[Population size effects in evolutionary dynamics]{Population
size effects in evolutionary dynamics on neutral
networks and toy landscapes}
\author{Sumedha$^1$,
Olivier C. Martin$^1,^2$, and Luca Peliti$^3$}
\address{$^1$ Université Paris-Sud, UMR8626, LPTMS, Orsay, F-91405; CNRS, Orsay,
F-91405, France}
\address{$^2$ UMR de Génétique Végétale du Moulon,
INRA-UPS-CNRS-INA PG, F--91190 Gif-sur-Yvette, France}
\address{$^3$ Dipartimento di Scienze Fisiche, Sezione INFN and Unità
CNISM, Università di Napoli ``Federico II'', Complesso
Universitario di Monte S. Angelo, I--80126 Napoli, Italy}
\eads{\mailto{sumedha.sumedha@gmail.com},
\mailto{olivier.martin@u-psud.fr},
\mailto{Luca.Peliti@na.infn.it}}
\begin{abstract}
We study the dynamics of a population subject to selective
pressures, evolving either on RNA neutral networks or in toy fitness
landscapes. We discuss the spread and the neutrality of the
population in the steady state. Different limits arise depending on
whether selection or random drift are dominant. In the presence of
strong drift we show that observables depend mainly on $M \mu$, $M$
being the population size and $\mu$ the mutation rate, while
corrections to this scaling go as $1/M$: such corrections can be
quite large in the presence of selection if there are barriers in
the fitness landscape. Also we find that the convergence to
the large $M \mu$ limit is linear in $1/M \mu$. Finally
we introduce a
protocol that minimizes drift; then
observables scale like $1/M$ rather than $1/(M\mu)$, allowing one to
determine the large $M$ limit faster when $\mu$ is small; furthermore
the genotypic diversity increases from $O(\ln M)$ to
$O(M)$.
\end{abstract}

\pacs{87.23.-n (Ecology and evolution),
87.15.Aa (Theory and modeling; computer simulation),
89.75.Hc (Networks and genealogical trees)}

\submitto{JSTAT}

Keywords: evolutionary dynamics, neutral networks, drift, selection

\maketitle

\section{Introduction}

In evolutionary biology, populations are subject to a number of
forces that shape their genetic composition~\cite{HerronFreeman03}.
Amongst these, mutations, selection and drift play a central role.
Drift becomes dominant for small populations, while for large
populations one reaches a steady state where mutations balance
effects of selection. The landscape
paradigm~\cite{Wright32,Gavrilets04} provides a relation between
genotype/phenotype and fitness, allowing for quantitative studies of
evolving populations, while at the same time giving a qualitative
picture. This has been particularly developed in the context of
quasispecies theory~\cite{Eigen71,EigenSchuster77,EigenMcCaskill89}
and of evolution on networks~\cite{Stadler02,Schuster02}; in a
different framework these problems have been analyzed within
evolutionary game
theory~\cite{LiebermanHauert05,TaylorRudenbert04,RocaGuesta06,AntalRedner06}
although, in that case, usually there is no drift in the usual sense and
the number of genotypes is low.

Let us consider a population of $M$ individuals evolving in a static
fitness landscape. We can define its steady-state distribution from
the average number of individuals with a given genotype, averaged
over a long stretch of time. One can also consider how
the population is spread out in genotype
space. If all genotypes have the same fitness (flat
landscape), the steady state distribution is independent of $M$
while the spread of the population depends mainly on the product $M
\mu$ where $\mu$ is the mutation rate~\cite{DerridaPeliti91}. In the
limit of an infinite population ($\mu$ fixed),
the generation-to-generation
fluctuations vanish and the instantaneous population distribution
coincides with the steady-state one. This is the quasispecies limit,
where the steady-state distribution is given by the leading
eigenvector of the evolution
operator~\cite{NimwegenCrutchfield99}. When the population
is finite, no general analytic solution for the steady-state
distribution is known. If the population size is much greater than
the number of genotypes, then the so-called diffusion
approximation~\cite{Ewens04} can be used. However, in most
realistic cases, the number of possible genotypes is much greater
than the population size. We are interested in understanding how
mutation, selection and drift affect different properties of the
steady state in such systems.

To unravel the different effects, we consider different evolutionary
dynamics, in which some of these processes may be present or not.
Most of our study is conducted in the framework of RNA neutral
networks~\cite{FontanaStadler93,Schuster02}, the archetypes of
genotype to phenotype mappings, but we also consider toy fitness
landscapes. After specifying our systems and population dynamics in
sections~\ref{sec:NeutralNetworks} to
\ref{sec:EvolutionaryDynamics}, we examine the dependence on the
mutation rate $\mu$ and on $M$ of the different processes driving
the dynamics. We begin with the case of neutral evolution in
section~\ref{sec:DynamicsWithoutSelection}, turning on and off the
drift. In section~\ref{sec:SelectionAndDrift} we allow for selection
in the usual way that leads to significant drift. As a general rule,
drift in these situations leads to $M \mu$ scaling, as shown
previously in the absence of selection~\cite{DerridaPeliti91}. Then
in section~\ref{sec:SelectionLowDrift} we introduce a particular
dynamics with selection but \emph{low drift}. There the $M \mu$
scaling is replaced by a smooth large $M$ limit even when $\mu \to
0$; furthermore the genotypic diversity becomes proportional to $M$
rather than being nearly constant. The corrections to these scalings
generically go as $O(1/M)$, with large effects when there are
barriers in the landscape, as we exhibit in
section~\ref{sec:Bottlenecks}. Section~\ref{sec:Conclusions} is
devoted to some final considerations.

\section{RNA neutral networks as fitness landscapes}
\label{sec:NeutralNetworks}

In studies of genotype to phenotype mappings, one often focuses on
biological molecules because the corresponding mapping is relatively
well defined. The genotype is simply the sequence of the
bio-molecule, while the phenotype is its shape, as specified for
instance by the minimum free-energy structure it folds into. Using
either protein or secondary RNA structures, it has been
found~\cite{Schuster02,FontanaStadler93,LipmanWilbur91,GrGiegerich96}
that neutral genotypes (genotypes that have a given phenotype) which
are connected via single mutational steps form extended networks
that permeate large regions of genotype space. These are known as
``neutral networks''. Via the neutral network, a population can move
in genotype space without crossing unfavorable low-fitness regions,
in contrast to what happens in many rugged fitness
landscapes~\cite{Wright32,KauffmanLevin87,
FlyvbjergHenrik92,NimwegenCrutchfield00}. However, because of the
huge dimensionality of our genotype space, large neutral (or nearly
neutral) networks can be argued to be inevitable~\cite{Gavrilets04}.

Here we shall work with an RNA neutral network, i.e., all RNA
sequences which fold into a given target RNA secondary structure.
The genotype of an RNA molecule is given by its base sequence: there
are four bases, A,C,G, and U, and thus $4^L$ genotypes for molecules
of $L$ bases. The molecule's phenotype is given by its
\emph{secondary} structure, i.e., by which bases are paired with which
as occur in its folded form. To every genotype one associates just
one phenotype (the secondary structure of minimum free energy) while
in general there will be many genotypes compatible with a given
phenotype. This many-to-one genotype to phenotype mapping has been
widely
studied~\cite{Schuster02,NimwegenCrutchfield99,FontanaStadler93,
GrGiegerich96,ReidysSchuster95,FontanaSchuster98,SchusterStadler02}.
Standard computational tools are available on the web to fold given
sequences; see for instance the \emph{fold} subroutine from the
Vienna package~\cite{Vienna95} which we used for all of this work's
computations. Two sequences are nearest neighbors (connected on the
neutral network) if and only if they differ by a single nucleotide
substitution. In general, RNA neutral networks are heterogenous
graphs, so that for instance the local connectivity varies quite a
lot from site to site.

\begin{figure}
\begin{center}
    \includegraphics[width=7cm]{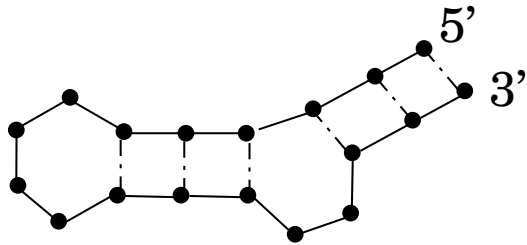}
    \end{center}
\caption{The target RNA secondary structure (cf.\ van Nimwegen et
al.~\cite{NimwegenCrutchfield99}).} \label{rna}
\end{figure}

The secondary structure (phenotype) chosen in this study is the one
used by van~Nimwegen et
al.~\cite{NimwegenCrutchfield99,NimwegenCrutchfield00}: it has $18$
nucleotides with six base pairs and is depicted in figure~\ref{rna}.
By single-nucleotide substitutions, purine-pyrimidine base pairs
(G--C, G--U, A--U) can mutate into each other, but not into
pyrimidine-purine (C--G, U--G, U--A) base pairs. Hence we considered
only the purine-pyrimidine base pairs. Given the base pairing rules
for this system, the number of a priori ``compatible'' sequences for
such a structure is $4^{6} \times 3^6 =2985884$. At $30^\circ$C,
37963 of these fold into the target structure; this number depends a
bit on the choice of temperature since the \textsl{fold} algorithm
computes free energies. We find these genotypes to be organized into
\emph{three} neutral networks (connected components), of sizes
$489$, $5784$ and $31484$ respectively. This will allow us to
investigate the effect of neutral network size on our observables.

To define the fitness landscape, we consider that individuals with
the ``correct'' phenotype (residing on the neutral network) are
viable, i.e., have maximal fitness, while all other phenotypes are
non-viable, i.e., have minimal fitness. We take these extreme values
to be 1 and 0, corresponding to the \emph{strong} selection limit;
then any mutation that takes one off the neutral network is lethal.

\section{Population spread, Hamming distances and neutrality}
\label{sec:neutrality}

We now ask how drift and selection affect observables associated
with the steady state population. A first observable quantifies how
much the population is ``spread out'' in genotype space, namely how
individuals at a given generation differ from one another in
genotype space. One defines the ``Hamming distance'' $h$ between two
genotypes as the number of positions where the two associated
sequences have different nucleotides. Following Derrida and
Peliti~\cite{DerridaPeliti91}, we shall study the distribution
$P(h)$ of distances when two genotypes are taken at random in the
population, averaged over generations. If $P(h)$ is broad, then the
population is spread out in genotype space.

A second observable is ``neutrality''~\cite{HuynenStadler96}. Let
$g_0$ be a genotype (a sequence of $L$ bases) belonging to the
neutral network; examine its $3 L$ possible single-nucleotide
substitutions and let $d$ be the number of these mutants that belong
to the neutral network. The ``neutrality'' of $g_0$ is then $d$, the
coordination (degree) of $g_0$ on the neutral network. A related
notion is the \emph{mutational robustness} $R_{\mu}$ of $g_0$. It is
defined as the survival probability of its mutant offspring. In the
context of neutral networks with fitness values 0 and 1, we see that
in fact
\begin{equation}
\label{eq:Rmu}
R_{\mu} = \frac{d}{3L},
\end{equation}
where $d$ is the neutrality of $g_0$. These definitions can be
straightforwardly extended to the neutrality or robustness of any
collection of genotypes. Thus one defines the ``network neutrality''
of a neutral network as the mean of $d$ when considering all of its
nodes. Similarly, when one has a \emph{population} of genotypes, the
``population neutrality'' is simply the average of $d$ over that
population, each individual being counted once. The population neutrality
depends on both the neutral network properties and on the
evolutionary dynamics~\cite{NimwegenCrutchfield99}. Furthermore, we
immediately see that the neutrality of a population is $3 L$ times
the mean robustness of its individuals.

\section{The model}
\label{sec:EvolutionaryDynamics}

Many processes affect the genetic makeup of natural populations. In
this work, we focus on the effects of mutation, selection and drift.
We wish to turn on and off selection or even drift while considering
the effects of the population size $M$ or of the mutation rate
$\mu$. Thus the details of the evolutionary process we consider are
tailored to emphasize one or another of these aspects at a time.

We consider a population evolving with nonoverlapping
generations; the population size is kept at a fixed value in the
standard way. Other choices could have been made, but as
in most studies, the detailed procedures used in the evolutionary
dynamics is not expected to be very important.

Given the $M$ individuals at the current generation, we must produce
$M$ viable offspring to form the next generation. Each offspring is
produced from a parent and given a chance to mutate: with
probability $1-\mu$, no mutation is applied, and with probability
$\mu$ one base at random is changed. Then \emph{selection} is
applied: the child is kept if and only if it is viable. More
generally, on an arbitrary fitness landscape, we let it survive
stochastically according to its fitness. Of course if there is no
selection, the offspring is always kept. The process of producing
offspring is repeated until the new generation has size $M$.
\emph{Drift} comes in via the way the parents are chosen to produce
candidate offspring. In the standard method, the parents are chosen
randomly \emph{with replacement}: clearly this allows for drift as
by bad luck some parents will not produce any offspring. In the
presence of selection, drift cannot be turned off completely but it
can be significantly lowered.

Indeed, let us consider the following process. First, each individual
of the population produces one offspring which mutates with
probability $\mu$: if a mutation is lethal, the corresponding
offspring is killed. In this step there is no replacement and the
resulting offspring population size will generally be smaller than
$M$. Second, one chooses individuals randomly from this offspring
population and replicates them. This is done until the population
size reaches $M$ again. Note that when $\mu$ is small, the new
generation will be nearly identical to the previous one, even for
small $M$, so there is very little drift. Because of selection, a
small amount of drift does occur, but its intensity is proportional
to $\mu$.

In all our runs, we initialize arbitrarily the population and let it
evolve for a large number of generations until initial conditions
are forgotten: this is the steady state limit. All the data
presented in this work are time averages taken from this regime. We
are now ready to see how $M$, $\mu$, drift and selection affect the
spread, neutrality and the genotypic diversity of the steady state
population.

\section{Dynamics without selection}
\label{sec:DynamicsWithoutSelection}

In this section, we consider a population evolving without being
subject to selection. We investigate the effects of allowing or not
drift, first on regular networks and then on
heterogeneous ones.

\subsection{Homogeneous networks}

Consider the space of sequences of length $L$; if we take these to
all be viable, then we get a homogeneous network in which all
1-mutant neighbors of each genotype belong to the network. Derrida
and Peliti~\cite{DerridaPeliti91} studied the evolution of a
population on such a flat fitness landscape. Using the fact that
there is no selection, it is easy to show that the steady state
distribution is \emph{uniform}. Thus the population neutrality is
trivial, being given by the degree of the network, i.e., $3L$, for
all population sizes. In contrast to this simple result, the
distribution of Hamming distances between genotypes in the
population is generally non-trivial. Depending on the nature of the
dynamics, we have the following behaviors:
\begin{itemize}
\item[(a)] Drift off --- The offspring are produced from parents
\emph{without replacement}: since each individual has exactly one
offspring, each lineage acts like an independent random walk. Thus
the Hamming distances between the genotypes in the population are
completely random: the mean of $h$ lies at $3L/4$ (at each position
along the sequence, one has a $3/4$ chance of having different bases
when comparing two random sequences) and its variance is equal to
$3L/16$.
\item[(b)] Drift on --- Here the offspring are produced from parents
\emph{with replacement}: the number of offspring of an individual is
variable, leading to tree genealogies. This situation incorporating
drift was studied by Derrida and Peliti~\cite{DerridaPeliti91} and
leads to a non-trivial $P(h)$ which depends on $M$ and $\mu$. At any
given generation, the individuals have mutual distances that reflect
the fact that they descend from a common ancestor, giving rise to a
clustering of the population that fluctuates from one generation to
the next. For our purposes here, we focus on the
result~\cite{DerridaPeliti91} that the relevant parameter when $M$
is large is $M \mu$: in particular, $P(h)$ depends only on the
product $M \mu$ at large $M$, a property that we call $M
\mu$~scaling.
\end{itemize}

\subsection{RNA neutral networks}

We now consider a population evolving on an RNA neutral network,
defined as the subspace of sequences which fold (at $30^\circ$~C)
into the target secondary structure shown in figure~\ref{rna}. RNA
neutral networks are generally heterogenous. Evolutionary dynamics
without selection can be implemented by simply ``forbidding''
attempts to apply lethal mutations. There are two natural ways to do
this, referred to as blind and myopic ant dynamics~\cite{Hughes96}.
In myopic ant dynamics, also called adaptive random
walks~\cite{KauffmanLevin87,FlyvbjergHenrik92,Orr03}, an offspring
that mutates is forced to choose a single point mutation that is
non-lethal (all non-lethal choices are equiprobable). In blind ant
dynamics (also called gradient random walks), a point mutation is
chosen at random (lethal or not): if it is non-lethal, it is
accepted, while if it is lethal, it is refused and the offspring is
taken to be \emph{non-mutant}. Both the blind and myopic dynamics
can be implemented with or without drift, according to the method
sketched in section~\ref{sec:EvolutionaryDynamics}. Although these
dynamical processes may appear to be somewhat artificial, they do
provide solvable cases. Furthermore, van~Nimwegen et
al.~\cite{NimwegenCrutchfield99} have shown that in the limit of
small $M \mu$, the standard evolutionary dynamics converges to the
blind ant dynamics.

Consider first the case without drift, in which the sampling of the
parents takes place \emph{without replacement}. Then each lineage
performs an independent random walk on the whole neutral network. As
shown by van~Nimwegen et~al.~\cite{NimwegenCrutchfield99}, the
steady-state distribution for blind ants is uniform on the neutral
network, while for myopic ants the probability of being at a node of
the neutral network that has degree $d$ is a constant times $d$.
(Note that if $d$ is the same for all nodes as in regular networks,
we obtain the uniform distribution as expected.) Given the
steady-state distribution, the histogram $P(h)$ of Hamming distances
is determined from the fact that the lineages are independent. At
large $L$, one expects it to become peaked, neutral networks being
widely spread out in genotype space. Furthermore, the steady-state
distribution and the $P(h)$ are $M$ and $\mu$ independent.

\begin{figure}
\begin{center}
    \includegraphics[width=7cm]{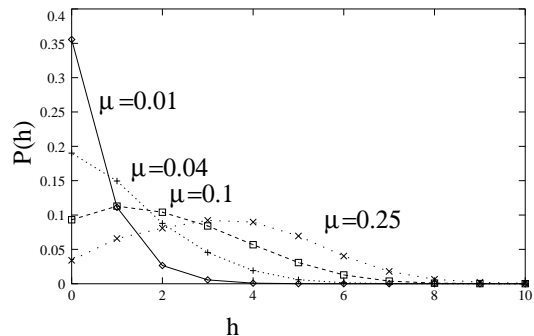}
\end{center}
\caption{Distribution of Hamming distances $h$ for different values
of $\mu$ for the myopic ant dynamics in the presence of drift. The
population size is $M=20$.} \label{hammingmy2}
\end{figure}

Let us now allow for drift. The sampling of the parents takes place
\emph{with replacement}. Interestingly, the steady-state
distribution of the population is not affected by the drift: this is
due to the fact that the heterogeneity of the neutral network does
not affect the chances of appearance of an offspring. However, the
lineages are no longer independent since the population typically
shares a recent common ancestor. As a consequence, the
Hamming-distance distribution $P(h)$ is not determined from the
steady-state distribution: it is non-trivial and depends
on $M\mu$ at large $M$. Since this result holds in a far
more general context which includes selection, we postpone its proof
to the next section. \emph{Corrections} to the $M \mu$ scaling are
$O(1/M)$, with typically a rather small prefactor. In the $M \mu \to
0$ limit, just as in the general population
dynamics~\cite{NimwegenCrutchfield99}, one recovers the blind ant
dynamics.

\begin{figure}
\begin{center}
    \includegraphics[width=7cm]{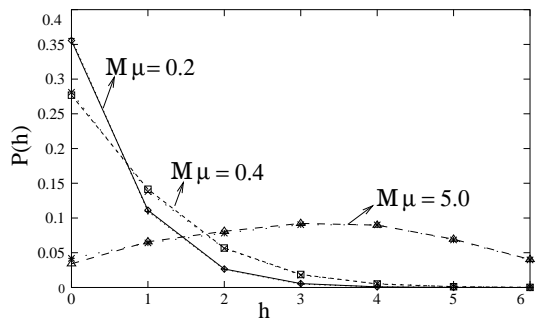}
\end{center}
\caption{Distribution of Hamming distances $h$ for different values
of $M \mu$ for the myopic ant dynamics in the presence of drift. For
each $M \mu$ we studied $M=20$ and $M=40$. The data with same $M
\mu$ superimpose perfectly, exhibiting the $M \mu$ scaling.}
\label{hammingmy1}
\end{figure}

For illustration, we show in Figures \ref{hammingmy2} and
\ref{hammingmy1} the distribution of Hamming distances $h$ for a
population with myopic ant moves in the presence of drift on a
neutral network. We see that in spite of the heterogeneity of the
network, the $M \mu$ scaling holds just as for a homogeneous network.

\section{Dynamics with selection and drift}
\label{sec:SelectionAndDrift}

\subsection{The infinite-population (quasi-species) limit}
\label{subsec:InfinitePop}

In the infinite-population limit, $M \to \infty$ (with $\mu$ fixed),
drift is absent and only the effects of mutation and
selection show up; this is called the quasi-species
regime~\cite{Eigen71,EigenSchuster77,EigenMcCaskill89}. As shown by
van~Nimwegen et~al.~\cite{NimwegenCrutchfield99}, the steady-state
distribution is given in this limit by the dominant eigenvector
$\Psi_0$ of a linear operator defined by the adjacency matrix of the
network. This eigenvector does not depend on $\mu$, but its
eigenvalue $\lambda_0$ does. If each individual in the population
produced only one offspring, and the unviable ones were eliminated
without replacement, then the population size would decay by a
factor $\lambda_0$ at each generation. Relating this decay to the
mutational robustness $R_{\mu}$ of the steady-state population we
immediately obtain
\begin{equation}
\label{eq:RmuInfPop}
\lambda_0 = (1-\mu) + \mu R_{\mu},
\end{equation}
which yields the population's average neutrality $\langle d
\rangle_{\infty}$ via equation~(\ref{eq:Rmu}): both $R_{\mu}$ and
$\langle d \rangle_{\infty}$ are $\mu$ independent.

These results can be compared to the population neutrality in the
case of blind/myopic ant moves, in which there is no explicit
fitness-based removal of individuals. In the case of blind ants, the
probability of residing on any node of the network is uniform. The
average neutrality seen by such a walker is just equal to the
average network neutrality. For a myopic ant, the probability of
choosing any node on the network is proportional to the degree of
the node. Neutrality is slightly higher in this case and is given by
the ratio of second and first moments of the node degrees. A
standard variational principle~\cite{SoshnikovSudakov03} shows that
the population neutrality is always at least as large as the network
neutrality, defined as the average degree of the neutral network.
Of course on a homogenous graph, population neutrality, blind ant
neutrality and myopic ant neutrality are all equal to the network
neutrality. We refer the reader to the work of van~Nimwegen
et~al.~\cite{NimwegenCrutchfield99} for a thorough discussion.


\subsection{Case of an infinite population at fixed $M \mu$}
\label{subsec:InfinitePopMmu}

In biological situations, the mutation rate is very small, making
the large-$M$ limit of little use: indeed to get to the infinite $M$
limit just discussed, the quantity $M \mu$ must be large.
When $M \mu$ is finite, drift occurs, and it is of interest
to consider the large $M$ limit at fixed $M \mu$. In
the \emph{absence} of selection, the $M \mu$ scaling law has been derived
by Derrida and Peliti~\cite{DerridaPeliti91}. In the presence of
selection, it has been exhibited from numerical simulations by
van~Nimwegen et~al.~\cite{NimwegenCrutchfield99}. It turns out
that this scaling follows
from the dynamical equations, whether or not there is selection:
these equations are invariant under a simultaneous rescaling of
time, $M$ and $\mu$, as long as $M \mu$ is fixed, as we now show.

Let $N_i(g)$ denote the number of individuals in the population
residing at the neutral network node $i$ at generation $g$. To go to
the next generation, let us choose $M$ individuals at random with
replacement, and let them mutate with probability $\mu$. One then
has
\begin{equation}
\label{eq:stochasticN}
N_i(g+1) = \sum_{\langle ij \rangle} P_j(M
p_j) + G_i(M q_i, M q_i),
\end{equation}
where $\sum_{\langle ij \rangle}$ denotes the sum over the nodes $j$
which are nearest neighbors of node $i$. Also, $p_j = \mu N_i / (3 L
M)$ is the probability to choose an individual of genotype $j$ and
to have it mutate to node $i$; $P_j(M p_j)$ is a Poisson random
variable of mean $M p_j$; $q_i = (1-\mu) N_i /M$ is the probability
to choose an individual of genotype $i$ and to leave it without
mutation; $G_i(M q_i, M q_i)$ is the sum of $M$ 0--1 random
variables of mean $q_i$, and thus at large $M$ is a Gaussian whose
mean and variance are both given by $M q_i$.

Since we are interested in the large $M$ limit with $M \mu$ fixed,
let us define $x_i =N_i/M$ to be the fraction of individuals
residing on node $i$. If we average equation~(\ref{eq:stochasticN}),
we recover
the deterministic (quasi-species) evolution equations for the
$x_i$'s. But fluctuations do \emph{not} go away at large $M$ if $M
\mu$ is fixed, instead the intensity of drift goes to a limit.
Extracting the
mean from the Gaussian of equation~(\ref{eq:stochasticN}), the
stochastic evolution equations for the $x_i$ take the form
\begin{equation}
\label{stochasticX} x_i(g+1) = (1-\mu) x_i(g) + \sum_{\langle ij
\rangle} P_j(M p_j) / M + G_i(0, M q_i) / M.
\end{equation}
Summing this expression over $M$ steps we obtain in
the limit of large $M$ (and thus $\mu \to 0$):
\begin{equation}
\label{stochasticXX}
\Delta x_i \equiv x_i(g+M) - x_i(g) = - M \mu x_i +
\sum_{\langle ij \rangle} (M \mu) x_j/3L + G_i(0,x_i),
\end{equation}
where we have used the fact that the sum of the $M$ Poisson
variables contributes via its mean but its variance (equal to $p_j$)
becomes negligible. Clearly, the steady-state behavior of these
stochastic equations depends only on the scaling parameter $M \mu$.
As expected, by averaging these equations, we recover the
quasi-species evolution dynamics. Strictly speaking, we have shown
that there is a limit when evolving $M$ individuals, neglecting the
decrease in the population size. But restoring the population to a
fixed size $M$ involves no mutations and so falls into the standard
case of drift for a single genotype; that stochastic process also
reaches a limit at large $M$ when $M \mu$ is fixed, so we can
conclude that the full process (which maintains the population size
at $M$) also depends only on the scaling variable $M \mu$ as $M \to
\infty$. Note that within this scaling framework, we recover the
$\mu$ fixed, $M\to \infty$ case of equation~(\ref{eq:RmuInfPop}) by
taking $M \mu \to \infty$: we shall see that the corrections to this
limit are linear in $1/M \mu$. One can also consider the limit $M
\mu \to 0$: there, the population structure typically collapses to
just one genotype at a time, and as shown by van Nimwegen et
al.~\cite{NimwegenCrutchfield99}, the effective dynamics reduces to a
random walk on the neutral network, so the population neutrality is
given by the network neutrality.

\subsection{Hamming distances in a finite population}
\label{subsec:HammingSelection}

We studied the distribution of Hamming distances between individuals
in the steady state population evolving on our three RNA neutral
networks. Here we report our results only for the largest network
(of size 31484), as qualitatively similar results were obtained with
the two other sizes.

\begin{figure}
\begin{center}
\includegraphics[width=7cm]{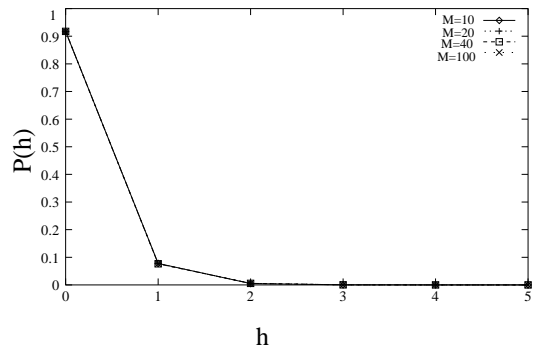}
\end{center}
\caption{Distribution of Hamming distances at $M \mu=0.2$ when
$M=10,20,40$ and 100. The data superpose perfectly.}
\label{rdsmmu0p2}
\end{figure}
\begin{figure}
\begin{center}
\includegraphics[width=7cm]{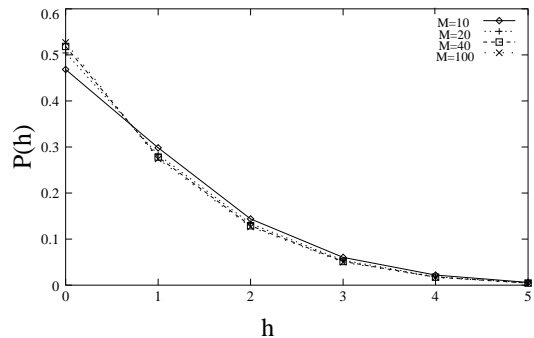}
\end{center}
\caption{Distribution of Hamming distances at $M \mu=2$ when
$M=10,20,40$ and 100. The data collapse is excellent when $M \ge 20$.}
\label{rdsmmu2}
\end{figure}
\begin{figure}
\begin{center}
\includegraphics[width=7cm]{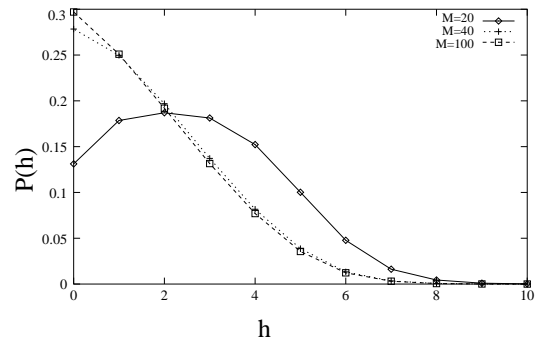}
\end{center}
\caption{Distribution of Hamming distances at $M \mu=5$ when
$M=20,40$ and 100. Here the data collapse requires $M \ge 40$.}
\label{hammingmmu5}
\end{figure}
\begin{figure}
\begin{center}
\includegraphics[width=7cm]{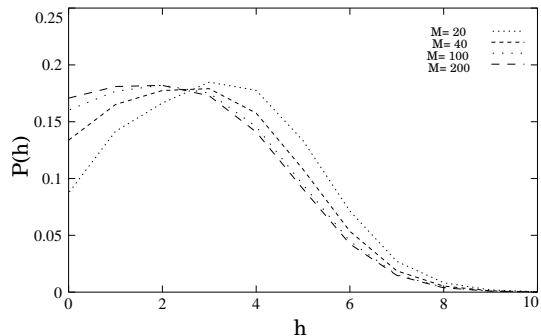}
\end{center}
\caption{Distribution of Hamming distances at $M \mu=10$ when
$M=20,40,100,200$. The $M \mu$ scaling is good only for $M \ge
100$.} \label{hammingmmu10}
\end{figure}

If we fix $M \mu$, we obtain $M$-independent results when $M$ is
large, in agreement with the scaling law derived in the previous
section. However, the value of $M$ at
which this scaling arises depends on $M \mu$. As shown in figures
\ref{rdsmmu0p2}, \ref{rdsmmu2}, \ref{hammingmmu5} and
\ref{hammingmmu10}, the scaling sets in for larger and larger values
of $M$ as the value of $M \mu$ grows. For instance, when $M \mu=10$,
one needs $M \ge 100$ to really see the $M \mu$ scaling convincingly
(cf.~figure~\ref{hammingmmu10}). Moreover corrections to this
scaling go as $O(1/M)$, i.e., they are of the same form as we found
in section~\ref{sec:DynamicsWithoutSelection} in the absence of
selection. This is a generic property that will be further studied
in section~\ref{sec:Bottlenecks}.

Within the $M \mu$ scaling, we see the population spread increases
monotonically with $M \mu$. In particular, as $M \mu \to 0$, the
spread goes to $0$, while as $M \mu \to \infty$, the population
spreads across the whole neutral network.

\subsection{Neutrality in a finite population}

We first examine the population neutrality $\langle d
\rangle_\mathrm{M}$. We are interested in seeing how large $M$ should
be for the $M \mu$ scaling to set in, considering in particular
the dependence on the neutral network size.

\subsubsection{Small neutral network}

\begin{figure}[htb]
\begin{center}
\begin{tabular}{cc}
\includegraphics[width=7cm]{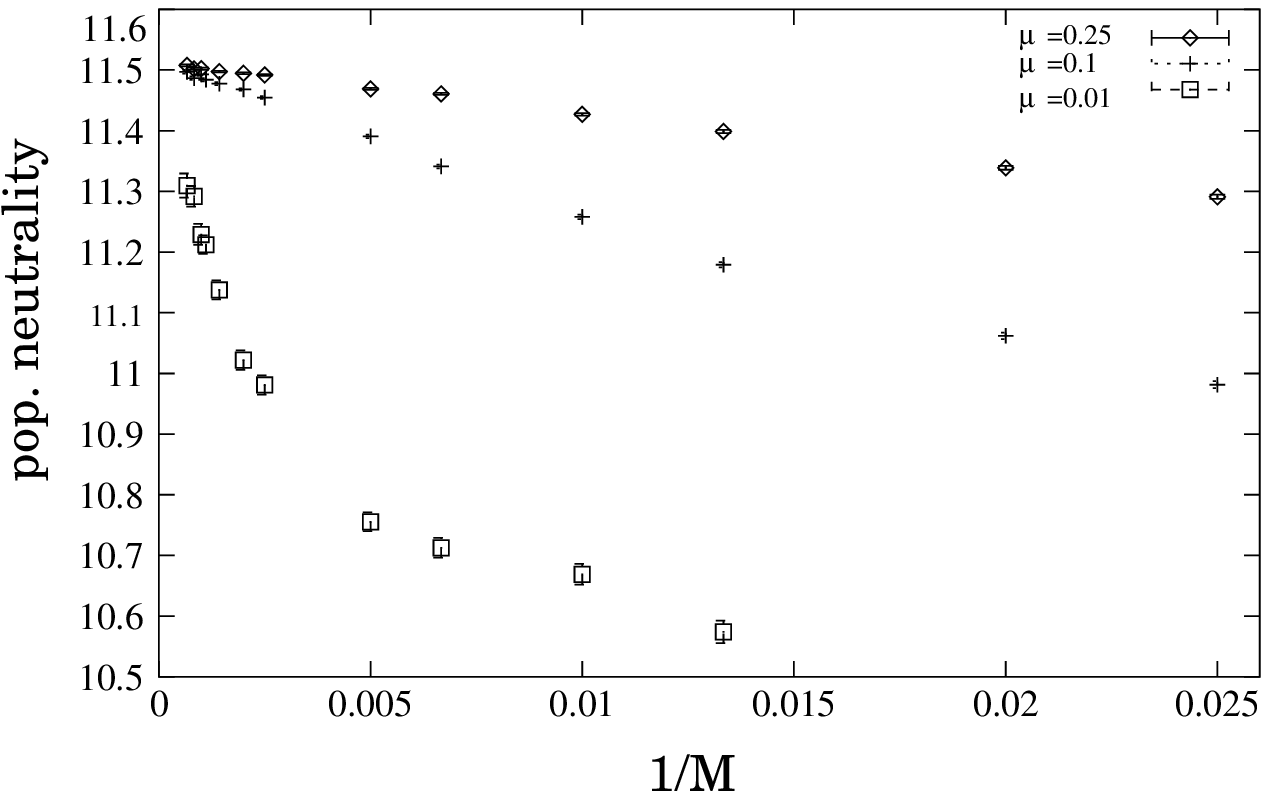}
&  \includegraphics[width=7cm]{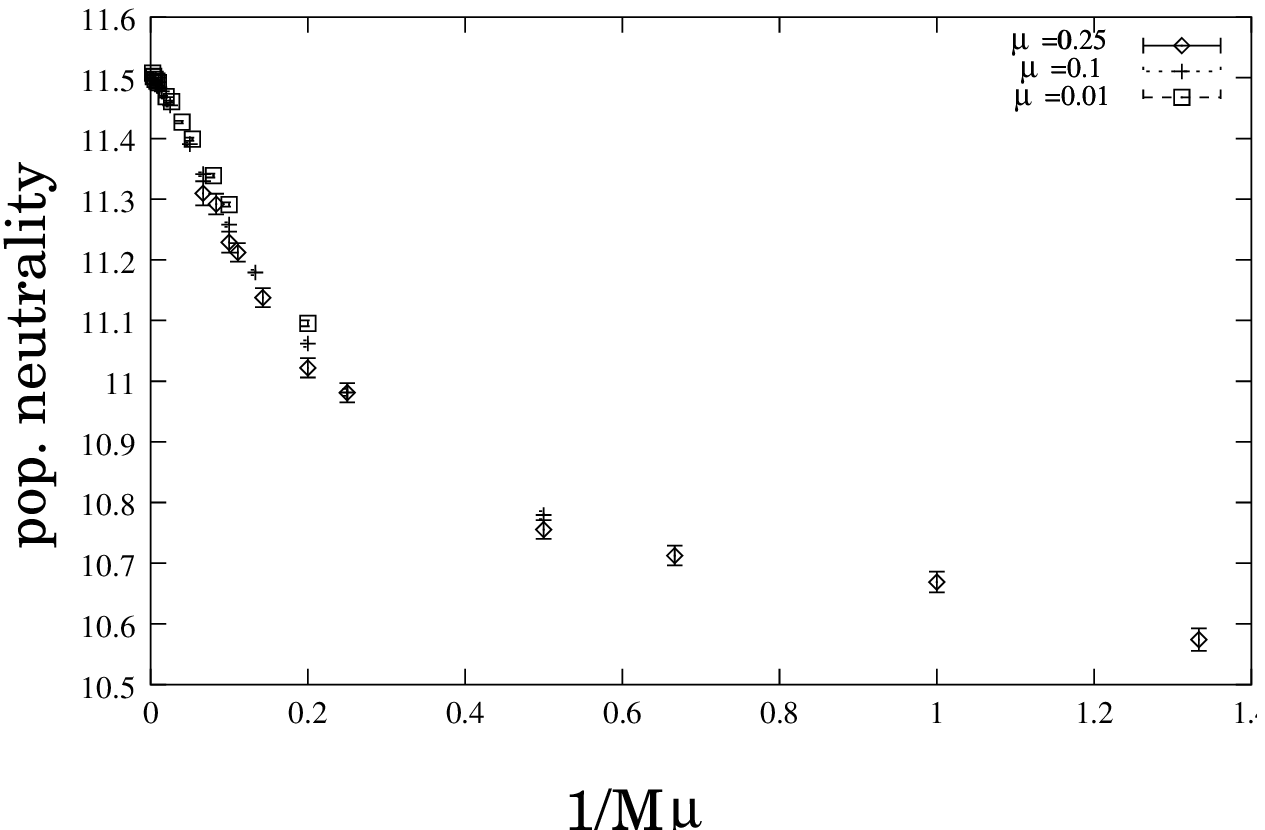}\\
\end{tabular}
\end{center}
\caption{Population neutrality $\average{d}_\mathrm{M}$ as a
function of $1/M$ and $1/M \mu$ for $\mu=0.01,0.1$ and $0.25$
in our small size network. One can
see the $M \mu$ scaling with small corrections due to
finite $M$; note also the linear behavior at the origin.} \label{small1}
\end{figure}

Figures~\ref{small1}(a) and (b) show the average neutrality
$\average{d}_\mathrm{M}$ as a function of $M$ and $M \mu$
respectively, for $\mu=0.01$, $0.1$ and $0.25$.

\begin{figure}
\begin{center}
    \includegraphics[width=7cm]{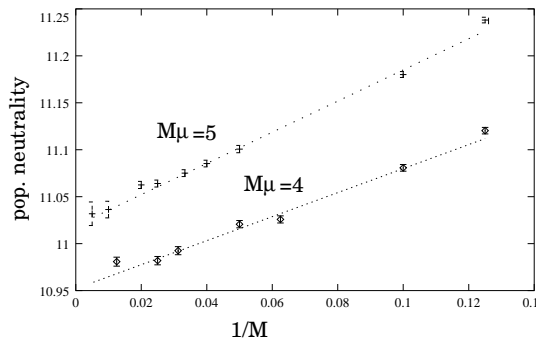}
\end{center}
\caption{$1/M$ law for the corrections to $M \mu$ scaling.} \label{small3}
\end{figure}

The average population neutrality depends both on $M$ and $M \mu$.
If $\mu$ is fixed and we take $M \to \infty$, we recover the
quasi-species limit which is
$\mu$-independent. Similarly, for fixed $M \mu$, we
obtain the $M \mu$ scaling regime by taking large $M$, just as in
van~Nimwegen et~al.~\cite{NimwegenCrutchfield99}. Furthermore, we
see that the corrections to the large-$M \mu$ limit of population
neutrality are linear in  $1/(M \mu)$ (cf.~the linear behavior at
the origin in Figure~\ref{small1}(b)). Finally, the approach to the
large $M$ limit at fixed $M \mu$ has measurable $1/M$ corrections.
(We also find that the value of $M \mu$ affects the time taken to
approach the steady state.) At fixed $M \mu$, the dependence on $M$
of the population neutrality or of the distribution of $d$ is rather
mild, though more marked than in the myopic or blind ant dynamics.
For several values of $M \mu$, we show these $1/M$ corrections in
figure~\ref{small3}. From all these data, we conclude that the
population neutrality has the form
\begin{equation}
\average{d}_\mathrm{M} = f(M \mu) \left(1+\frac{A(M
\mu)}{M} + \cdots\right),
\label{eq:CorrectionsToScaling}
\end{equation}
where $f(M \mu)=\average{d}_{\infty}$ is the $M \rightarrow \infty$
limit of $\average{d}_\mathrm{M}$ at given $M \mu$ and
\begin{equation}
f(M \mu) = f(\infty) \left(1 + \frac{B}{M \mu} + \cdots \right),
\label{eq:CorrectionsToInfinity}
\end{equation}
describes how the large $M \mu$ limit converges to the quasi-species
case. Finally, as shown by van~Nimwegen
et~al.~\cite{NimwegenCrutchfield99}, $f(M \mu)$
tends to the network neutrality as $M \mu \to 0$.

\subsubsection{Medium and large networks}

\begin{figure}[htb]
\begin{center}
\begin{tabular}{cc}
\includegraphics[width=7cm]{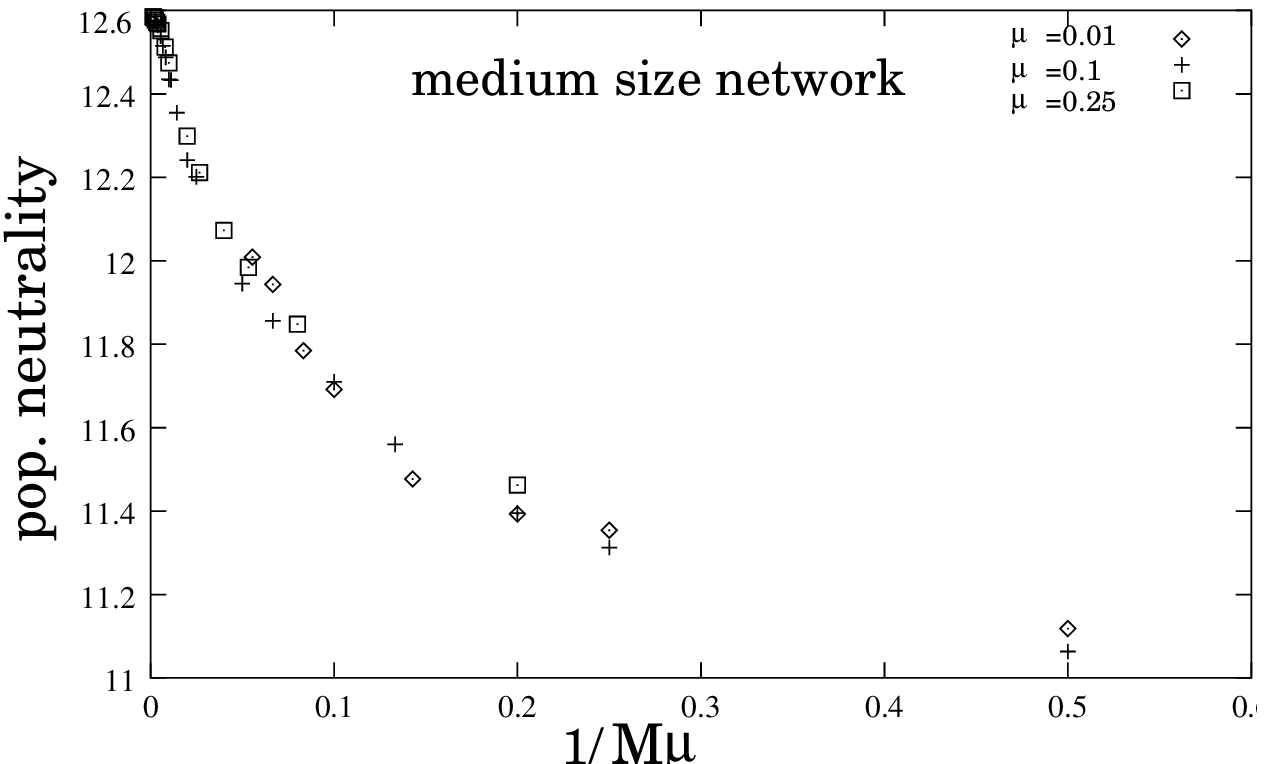} &
    \includegraphics[width=7cm]{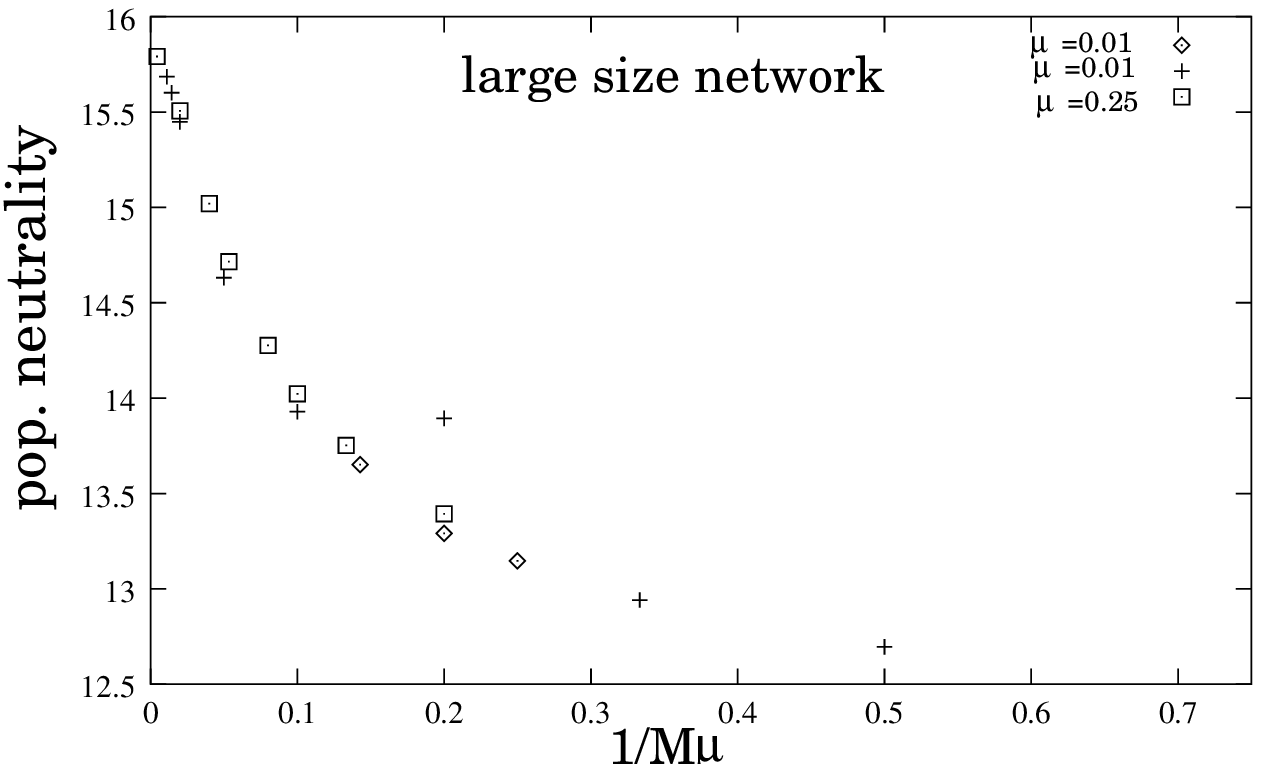} \\
\end{tabular}
\end{center}
\caption{Average neutrality as a function of $1/M \mu$ for different
values of $\mu$ on our medium and large size networks.} \label{med}
\end{figure}

Similar results are found for our medium and large networks.
Figure~\ref{med} shows the population neutrality
as a function of $1/M \mu$,
exhibiting good $M \mu$ scaling.
However the corrections to this scaling are larger than those
found in the small network; in fact, to see the asymptotic $1/M$ law
for the corrections to the $M \mu$ scaling, one has to go to quite
large values of $M$. Finally, as before, the
quasi-species limit is reached with corrections $O(1/(M \mu))$.

Other results that seems to apply generally is that for fixed $\mu$,
the population neutrality (and thus the population robustness)
increases with $M$, but it decreases at fixed $M \mu$.
Mathematically, this means that $A > 0$ in
equation~(\ref{eq:CorrectionsToScaling}) while $B < 0$ in
equation~(\ref{eq:CorrectionsToInfinity}).

\section{Dynamics with selection but low drift}
\label{sec:SelectionLowDrift}

\subsection{Framework}
\label{subsec:framework}

Random drift reduces the population spread and thus delays the
approach to the large $M$ limit at fixed $\mu$. Lowering
the drift
would thus allow one to reach the large $M$ limit more easily.
Furthermore, one would have a higher mutational robustness of the
steady-state population for a given population size; this higher
survival probability suggests that biological mechanisms for
reducing drift~\cite{WhitlockIngvarsson00} could be selected for in
natural populations. In this section we study dynamics in which
selective pressures are high but the drift is particularly low.

Our dynamics on a neutral network is defined as follows
so that drift effects are minimized:
\begin{itemize}
\item For a given population of size $M$ and mutation probability
per individual $\mu$, we have $M \mu$ of the individuals in the
population hop to any of their $3L$ neighbors; the others are
unchanged. If a mutation brings an individual off the neutral
network, we kill it, otherwise we keep it.
\item We find the number of individuals that were killed and replace them
by randomly cloning individuals from the remaining population.
\end{itemize}
Note that in these dynamics we perform a random sampling \emph{only
for a part of the population}; in fact, in the absence of selection,
there is no drift at all.

To understand the essential difference from the usual dynamics, we
use a branching process representation of the evolution and the
cloning. An individual is represented by a point; at a given time
there are $M$ points for $M$ individuals (note that as always $M$
represents the number of individuals, not the number of different
genotypes). From one generation (time) to the next,
$M \mu$ points are replicated leading to branchings
while the other points just proceed without branching. The time
evolution generates branching processes in which each individual is
represented by a point and is connected to its parent at the
previous time by an edge. Following this branching process, we
obtain the descendants of a given individual; one can also
consider two individuals, follow their edges backwards till one reaches
their most recent common ancestor. Because of the drift, at large
enough times the whole population belongs to just one connected
component.

Following Derrida and Peliti~\cite{DerridaPeliti91}, let us
investigate the genealogy of individuals going back in generations.
For a subset of $k$ individuals in the population, let $w_k(t)$ be
the probability that their ancestors $t$ generations ago were all
different. To calculate this quantity, we first determine the
probability $x_k$ that $k$ individuals have $k$ distinct immediate
ancestors (parents). Define $q=\mu(1-\average{R_\mu}_{M})$ where
$\average{R_\mu}_{M}$ is the mean robustness of the population. Then
assuming $q$ is small and neglecting node to node fluctuations in
neutrality, we have
\begin{equation}
x_k \approx 1 - \frac{k (k-1) q}{(1-q)M} \ .
\end{equation}
Taking the generation-to-generation processes to be independent, one
has $w_k(t) = x_k^t$, which can be approximated to leading order in
$q$ by
\begin{equation}
\label{DP2:eq}
    w_k(t) \approx \exp\left(-\frac{k(k-1) q}{M} ~ t\right).
\end{equation}
Hence, unlike the random drift case, the time scale is proportional
to $M/q$. We thus define a rescaled dimensionless time variable
\begin{equation}
    \tau = q t / M.
\end{equation}
The probability that the two individuals shared a common ancestor at
most $t$ generations back is given by $1-w_2(t) = 1-\exp(-2\tau)$.
Thus the probability density $p(\tau)$ that the most recent common
ancestor of the two individuals arose between $\tau M/q$ and
$(\tau+d \tau) M/q$ generations ago is given by
\begin{equation}
    p(\tau)=\frac{\rmd (1-w_2(\tau))}{\rmd \tau} = 2 \exp(-2\tau).
\end{equation}
Therefore the characteristic time for the most recent common
ancestor scales as $M/\mu$, while it scales as $M$ in the usual
dynamics with drift of section~\ref{sec:SelectionAndDrift}.

The distribution of times to the most recent
common ancestor can be used to
obtain the Hamming-distance distribution of the steady-state
population. Let $\phi_{\nu}(t)$ be the probability that two random
walkers in genotype space find themselves at a Hamming distance
$\nu$ given that they coincided $t$ generations before. We have
\begin{equation}
\label{DP3:eq}
\phi_{\nu}(t)= \frac{\Gamma[L+1]}{2^{\nu} \Gamma[\nu+1]
\Gamma[L-\nu+1]} (1-\exp^{-2 \mu t})^{\nu},
\end{equation}
where $L$ is the genome length. Then $P_{\nu}$, the probability that
the Hamming distance between two individuals in the steady state
population is equal to $\nu$, is given by
\begin{equation}
    P_{\nu} = \int_{0}^{\infty} \rmd \tau\, p(\tau) \phi_{\nu}(t).
\end{equation}
Using the expression of $\phi_{\nu}(t)$ from equation~(\ref{DP3:eq}),
one sees that $P_{\nu}$ depends on $M$ but not much
on $\mu$; in particular, one has
a well defined $\mu \to 0$ limit at fixed $M$. This is to be
contrasted with the random drift case where $P_{\nu}$ depends mainly
on $M \mu$.

\subsection{Consequence for the genotypic diversity}
\label{subsec:diversity}

A high level of genotypic diversity in a population is usually
advantageous for survival. Here we study how low drift can greatly
enhance this diversity by examining two measures of the number of
different genotypes, namely the actual number $g_M$ (which is
frequency independent), and the inverse participation ratio $G_M$ of
the genotypic abundances. Explicitly, for a population of size $M$,
if the number of individuals with genotype $i$ is $m_i$, we define
\begin{equation}
G_M = \frac{(\sum m_i)^2}{\sum m_i^2},
\end{equation}
where the sum runs over all the different genotypes present in the
population.
\begin{figure}
\begin{center}
    \includegraphics[width=7cm]{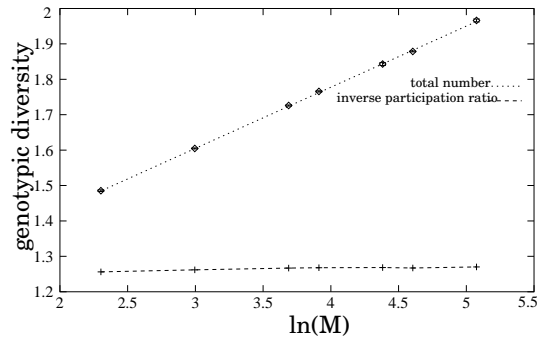}
\end{center}
\caption{Plot of genotypic diversity ($g_M$ and $G_M$) for dynamics with
  selection and drift as
function of $\ln M$ at $M \mu =0.4$.}
\label{fig:driftd}
\end{figure}
\begin{figure}
\begin{center}
    \includegraphics[width=7cm]{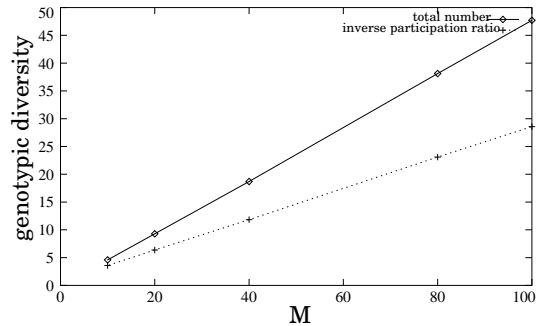}
\end{center}
\caption{Plot of genotypic diversity ($g_M$ and $G_M$)
as a function of population size
$M$ for dynamics with selection but low drift at $M \mu =0.4$.}
\label{fig:lessdriftd}
\end{figure}

We show in figure~\ref{fig:driftd} the two different measures of
genotypic diversity for dynamics with selection and drift. The
bottom curve corresponds to $G_M$, the top one to $g_M$. We find
that the ``absolute'' genotypic diversity $g_M$, (taking into
account genotypes of arbitrarily low frequencies), grows
logarithmically (and thus rather slowly) with $M$ at fixed $M \mu$.
On the other hand, the rare genotypes contribute less to the inverse
participation ration $G_M$, and we find that this measure of
genotypic diversity saturates in the large $M$ limit at fixed $M
\mu$.

In Fig.~\ref{fig:lessdriftd} we display the same quantities for our
low-drift dynamics. We now see that both measures of diversity grow
linearly with $M$ at fixed $M \mu$; thus each genotype in the
population arises just a few times as $M \to \infty$ if $M \mu$ is
fixed. Clearly the reduction of drift in this dynamics allows for a
high genotypic diversity.

\subsection{Hamming distances}
\begin{figure}
\begin{center}
    \includegraphics[width=7cm]{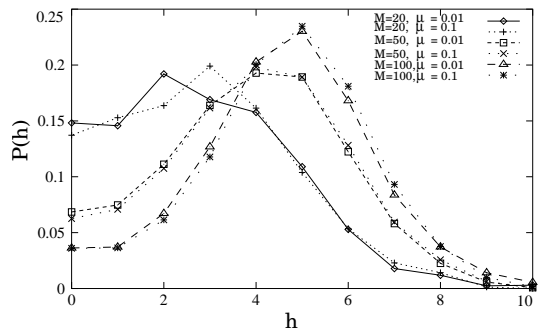}
\end{center}
\caption{Distribution of Hamming distances
for evolutionary dynamics with low drift. The distributions
depend strongly on $M$ and only very weakly on $\mu$.} \label{hammings1}
\end{figure}

\begin{figure}
\begin{center}
    \includegraphics[width=7cm]{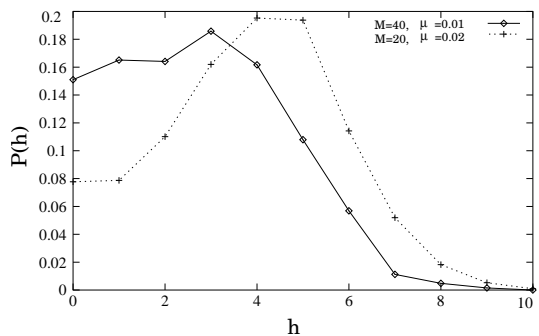}
\end{center}
\caption{Distribution of Hamming distances for two different values of
population size (20 and 40) when $M \mu$=0.4, showing the absence
of $M \mu$ scaling when the dynamics has low drift.}
\label{hammings2}
\end{figure}

We can study Hamming distances as was done in
section~\ref{subsec:HammingSelection} for our low-drift dynamics. We
find that the distribution of Hamming distances depends strongly on
$M$ but not much on $\mu$ and approaches a $\mu$-independent limit
as $M$ grows large (figures~\ref{hammings1} and~\ref{hammings2}).
This is consistent with the above calculation for the scaling laws.
As expected, for fixed $M$, there is a non-trivial limit
distribution as $\mu \to 0$.

\subsection{Population neutrality}

\begin{figure}[htb]
\begin{center}
\begin{tabular}{cc}
\includegraphics[width=7cm]{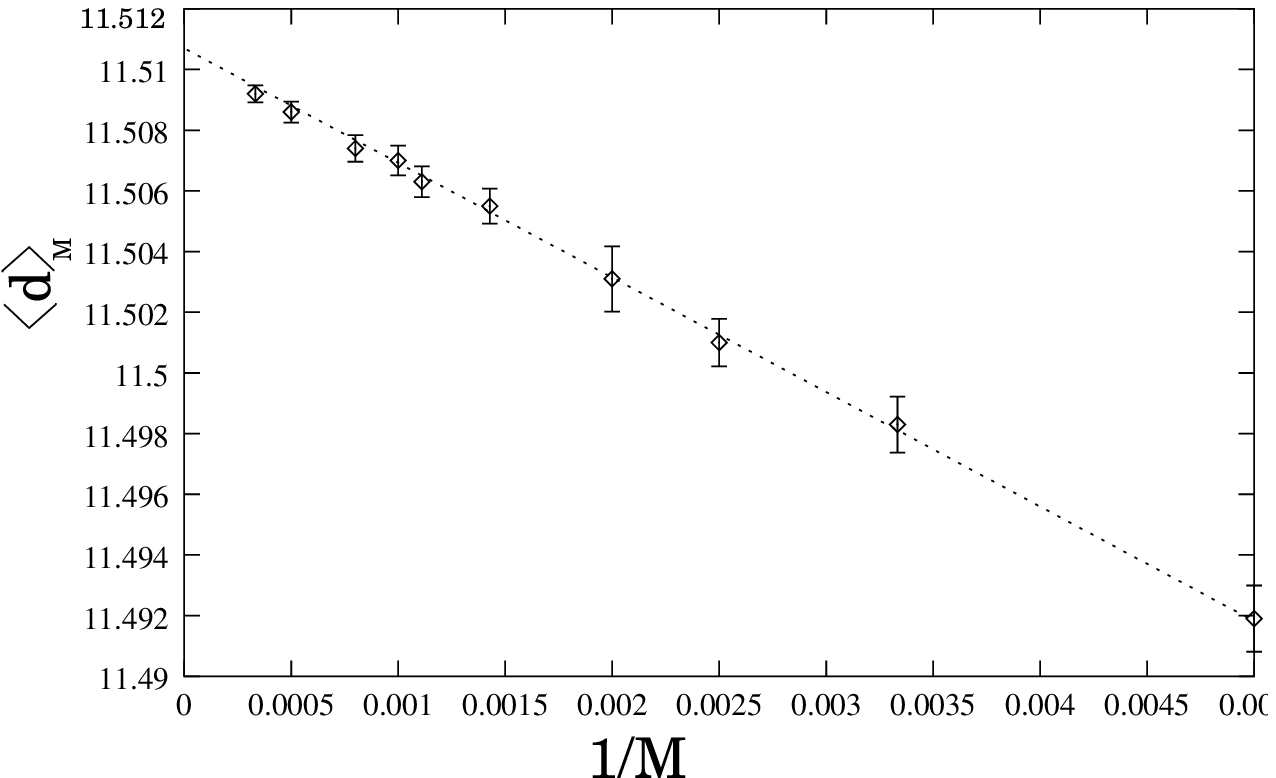} &
    \includegraphics[width=7cm]{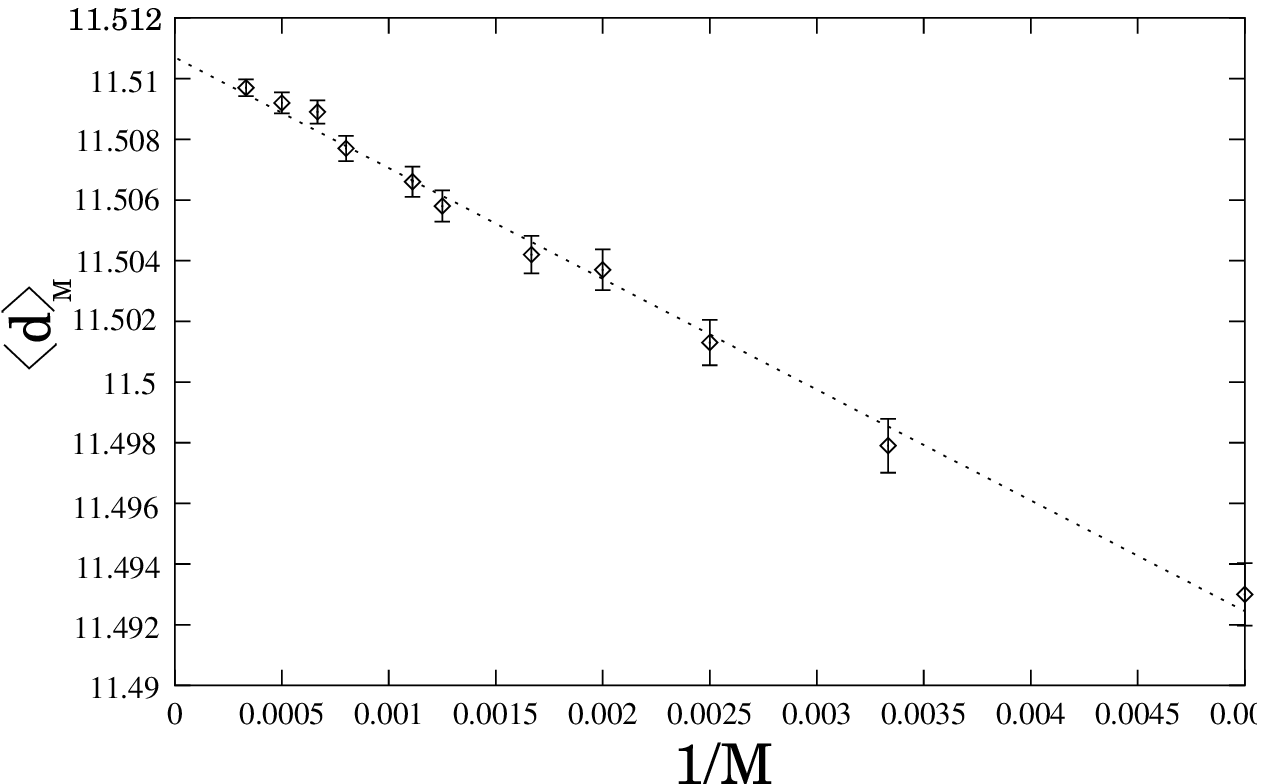} \\
\mbox{\bf ($\mu$ =0.1)} & \mbox{\bf ($\mu$ =0.25)}
\end{tabular}
\end{center}
\caption{Population neutrality vs $1/M$ on the small size network.
The dotted line is a linear fit as in Eq.~(\ref{eq:nc}).}
\label{small}
\end{figure}

In the infinite population limit, the value of population neutrality
is independent of drift effects, and it is also independent of
$\mu$. Consider first our small neutral network with $489$ nodes and
network neutrality $10.4499$; we determined the largest eigenvalue
of the adjacency matrix (via
\textsl{Mathematica}~\cite{Mathematica05}), obtaining $\langle d
\rangle_{\infty} = 11.5107$. In a finite population we find that
$\average{d}_M$ approaches $\average{d}_{\infty}$, with $1/M$
corrections that do not depend much on the mutation rate. In
particular, we find
\begin{equation}
    \average{d}_M = \average{d}_{\infty} \left(1+\frac{A(\mu)}{M} \right),
\label{eq:nc}
\end{equation}
where $A=-0.328\pm 0.002$ for $\mu =0.1$ and $A=-0.317 \pm 0.005$
for $\mu=0.25$. The corresponding fits are shown in
figure~\ref{small}.

\begin{figure}[htb]
\begin{center}
\begin{tabular}{cc}
\includegraphics[width=7cm]{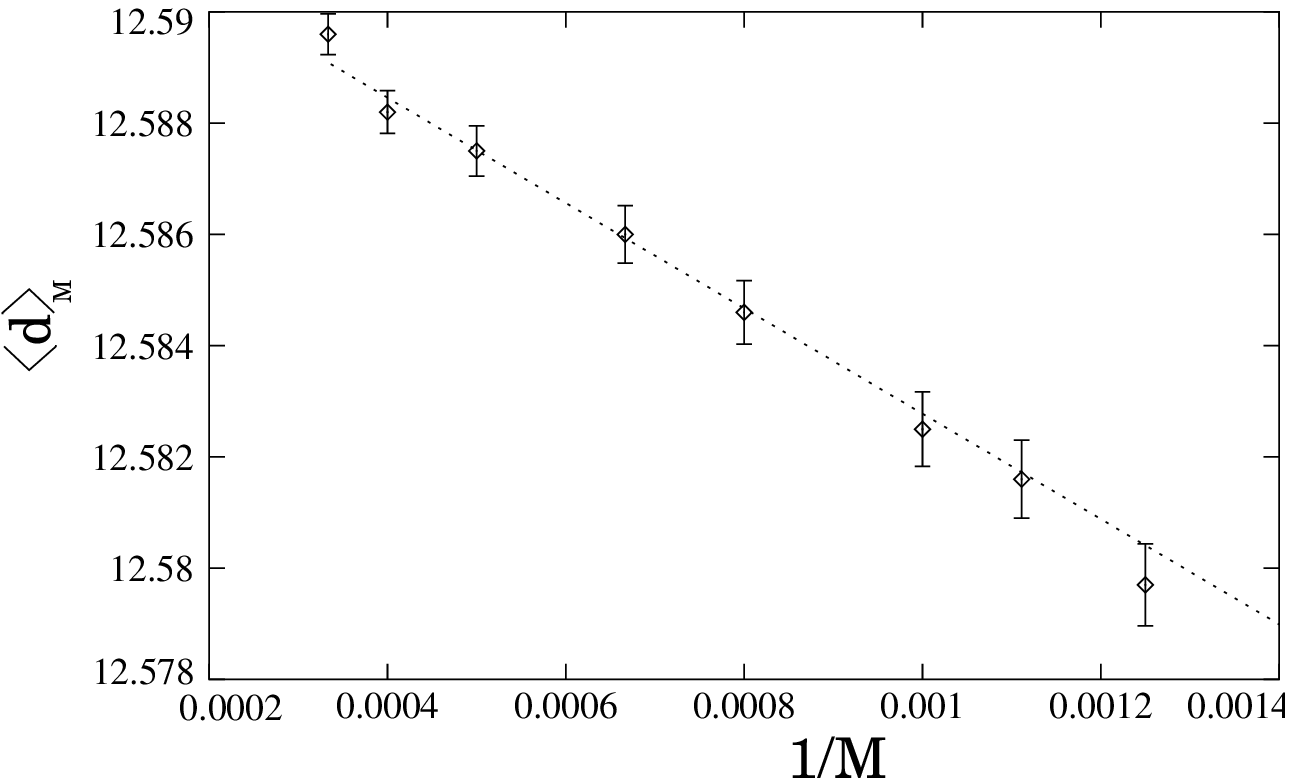} &
    \includegraphics[width=7cm]{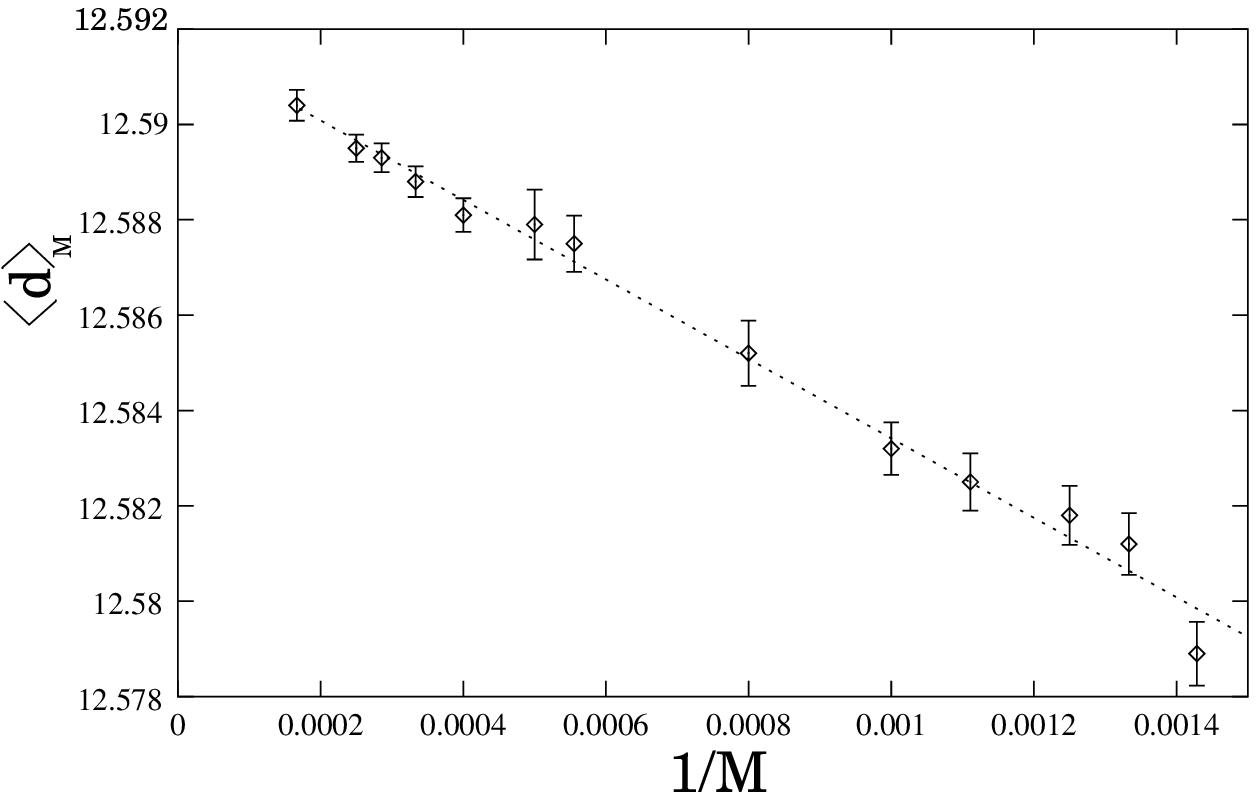} \\ [0.4cm]
\mbox{\bf ($\mu$ =0.05)} & \mbox{\bf ($\mu$ =0.25)}
\end{tabular}
\end{center}
\caption{Population neutrality vs $1/M$ on the medium size
network. The dotted line is a linear fit as in Eq.~(\ref{eq:nc}).}
\label{medium}
\end{figure}

On our medium-size neutral network (with $5784$ nodes), we find that
the network average neutrality is equal to $10.6888$, while
$\average{d}_{\infty} = 12.592 \pm 0.0002$. We again find a $1/M$
convergence as in Eq.~(\ref{eq:nc}) with $A=-0.752 \pm 0.035$ for
$\mu=0.05$ and $A= -0.662 \pm 0.02$ for $\mu=0.25$
(figure~\ref{medium}). Just as for the small neutral network, $A$
does not depend much on $\mu$.
\begin{figure}[htb]
\begin{center}
\begin{tabular}{cc}
\includegraphics[width=7cm]{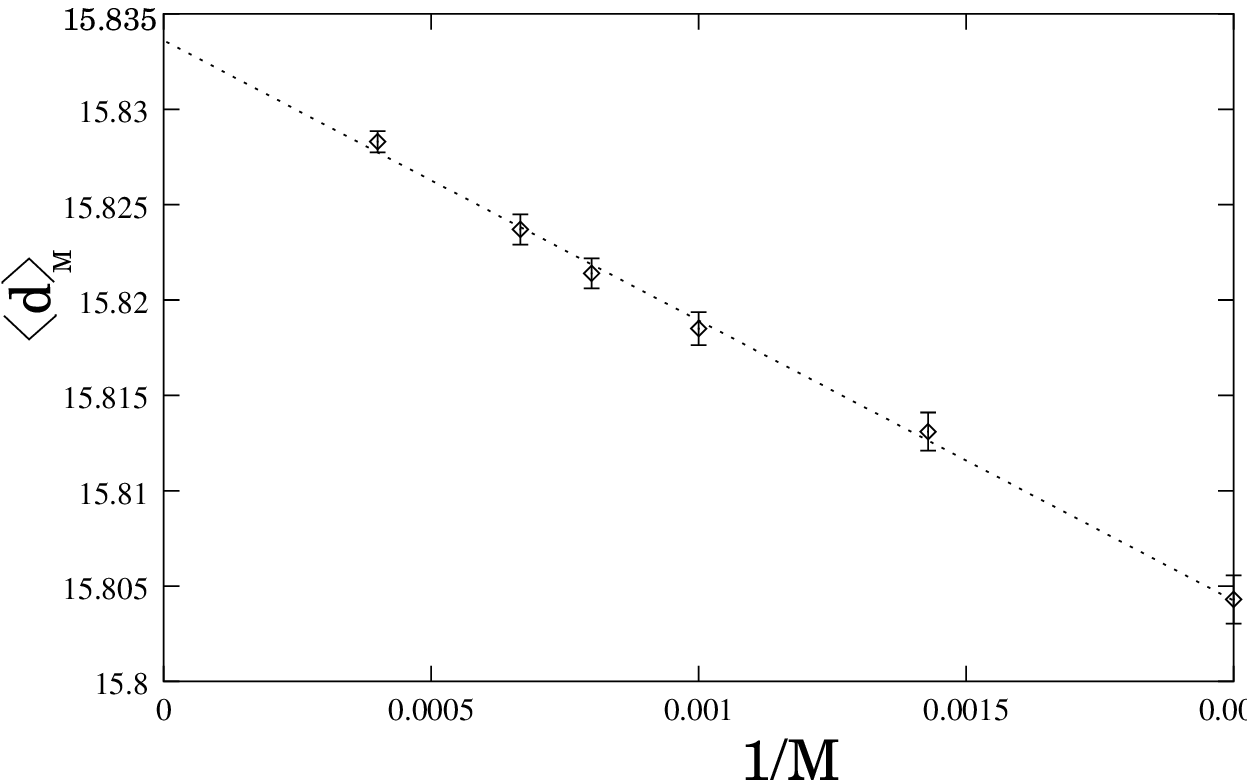} &
    \includegraphics[width=7cm]{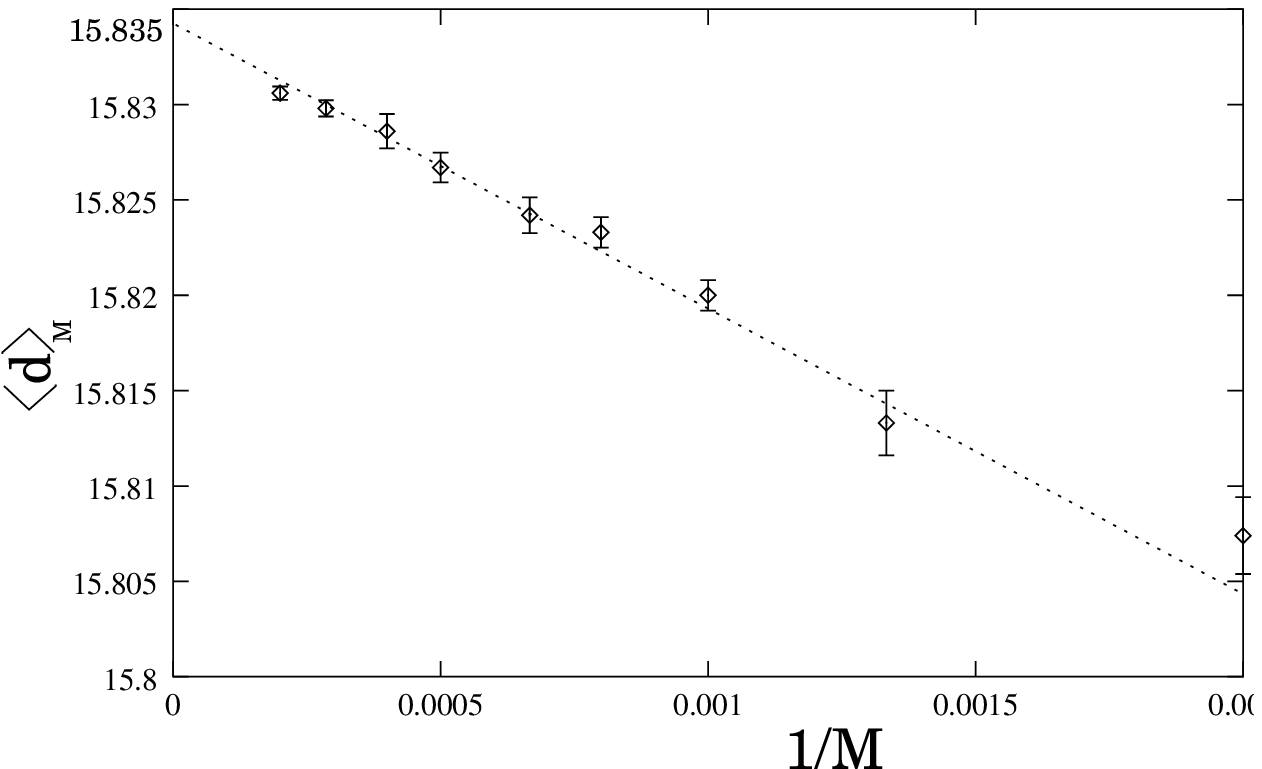} \\
\mbox{\bf ($\mu$ =0.05)} & \mbox{\bf ($\mu$ =0.25)}
\end{tabular}
\end{center}
 \caption{Population neutrality vs $1/M$ on the large size network.
The dotted line is a linear fit as in Eq.~(\ref{eq:nc}).}
\label{large}
\end{figure}
%

Finally, on the large neutral network (with $31484$ nodes), we find
the network neutrality to be $12.116$, whereas $\average{d}_{\infty}
= 15.434$. Figure~\ref{large} confirms the $1/M$ convergence with
$A=-0.927 \pm 0.045$ for $\mu =0.05$ and $A=-0.944 \pm 0.044$ for
$\mu =0.25$.

\begin{figure}
\begin{center}
\includegraphics[width=7cm]{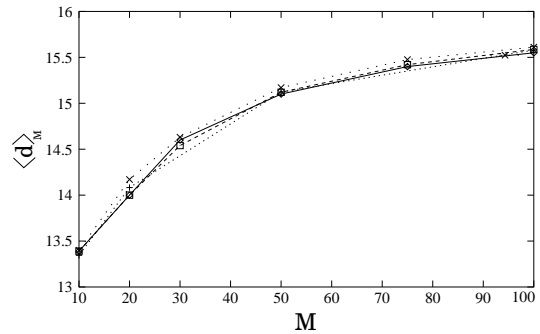}
\end{center}
\caption{Population neutrality $\average{d}_M$ vs $M$ for
$\mu =0.01,0.05,0.1$ and $0.25$ on the large neutral network,
showing the insensitivity to $\mu$.}
\label{Mdep}
\end{figure}

\begin{figure}
\begin{center}
\includegraphics[width=7cm]{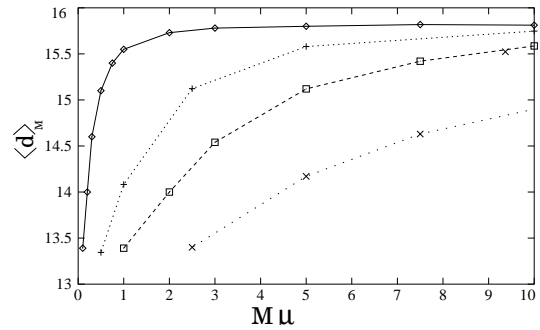}
\end{center}
\caption{Population neutrality $\average{d}_M$ vs $M \mu$ for $\mu
=0.01,0.05,0.1$ and $0.25$ on the large neutral network,
showing the absence of $M \mu$ scaling. }
\label{Mmudep}
\end{figure}

The overall pattern is thus that the data are well represented by
equation~(\ref{eq:nc}),with an $A(\mu)$ that grows with increasing
network size and depends slowly on $\mu$. We have also checked
directly that $\average{d}_M$ is rather insensitive to the value of
$\mu$ (cf.~figure~\ref{Mdep}). Furthermore, in all cases, there is
no $M \mu$ scaling (cf.~figure~\ref{Mmudep}), in contrast with what
happens for the case of standard drift (see
section~\ref{sec:SelectionAndDrift}).

\subsection{Distribution of neutrality}

\begin{figure}
\begin{center}
\includegraphics[width=7cm]{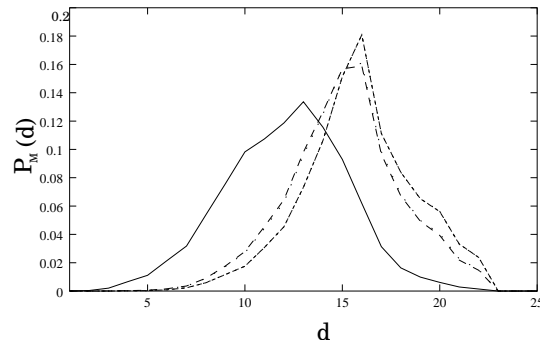}
\end{center}
\caption{The continuous line shows the distribution of
the number of neighbors for
random nodes on the largest network (with 31484 nodes). The curve closest
to it shows the steady state distribution for two data sets,
with $\mu=0.1$ and $0.25$ when the population size is 50:
they superpose extremely well.
The third curve corresponds to $M=1000$ and
$\mu=0.1,0.25$ and $0.5$, again with excellent superposition.
Superposition (independence on $\mu$) becomes
exact when $M=\infty$.} \label{dis}
\end{figure}

As a last point, let us consider the whole distribution $P_M(d)$
rather than just the average population neutrality $\average{d}_M$.
In figure~\ref{dis} we display several cases of interest for our
largest neutral network. The left-most curve is for genotypes chosen
randomly and uniformly from the neutral network. The next curve on
the right is for a population of size $M=50$ undergoing selection
with low drift; there are in fact two data sets displayed, one for
$\mu=0.1$ and one for $\mu=0.25$. The last curve is for the same
algorithm at $M=1000$ for three values of $\mu$, namely $\mu=0.1$,
$0.25$ and $0.5$.

Several comments are in order. First, as $M$ increases, the overall
trend is for $P_M(d)$ to shift to larger $d$; this is in agreement
with the general property that mutational robustness grows with
increasing genotypic diversity. Second, there is hardly any
dependence of these data on $\mu$, a feature particular
to dynamics having reduced drift.

\section{Reaching the large $M$ limit and barriers in the fitness landscape}
\label{sec:Bottlenecks}

\subsection{Motivation}
For large populations in the presence of drift, the relevant
parameter appears to be $M \mu$, as we have seen in
sections~\ref{sec:DynamicsWithoutSelection} and
\ref{sec:SelectionAndDrift}. When one wishes to evaluate the
large-$M$ limit at given $\mu$, it is better to minimize drift as in
section~\ref{sec:SelectionLowDrift}, especially when $\mu$ is small,
since $\mu$ plays little role in this modified dynamics. We also saw
that the corrections to the $M=\infty$ limit go like $O(1/M)$ on
neutral networks. Simple arguments~\cite{CerfMartin95} suggest that
this is a generic property of growth with
diffusion phenomena, and we shall now confirm this on toy
landscapes: the $O(1/M)$ corrections are not an artefact of the
fitness being 0 and 1 as we have assumed so far. Furthermore, we
wish to get some insight into the \emph{size} of this correction.
For our three neutral networks, we found that the factor $A$ of the
$A/M$ correction grows with increasing neutral network size;
however, barriers (entropic or fitness) in the landscape are likely
to affect $A$, as we will now illustrate using a few toy landscapes.

\subsection{Evolution in a toy fitness landscape}

To build up our intuition, we will consider a space where genotypes
are parametrized by a real number, and mutations correspond to small
changes of this number. Evolution in low-dimensional fitness
landscapes has been considered by several authors in the recent
years and has led to a number of insights (see, e.g.,
\cite{tsimring96,tsimring97,peng03,rouzine03}). In this case, it is
convenient to consider a continuous-time limit because this allows
for a Fokker-Planck formulation. We can start with all individuals
at the same position or place them randomly in the landscape. After
some time, the population reaches a well defined steady state. For
the numerical simulation of such evolutionary dynamics, we
discretize time using a time step $\Delta t$; this then
gives the following update rules:
\begin{itemize}
    \item At each
step (small) mutations arise; this means that the genotype $x$ is
changed by $\Delta x$ where $\Delta x$ is a Gaussian random
variable of standard deviation $\sqrt{2 D \Delta t}$~\cite{Vankampen81}.
    \item Given the new positions of the individuals, we allow for
replication according to fitness given by a function
$-V(x)$ in the Fokker-Planck language.
$V(x)$ is low (or even negative) if
the genotype has a high fitness; it is large and positive for an
unfit genotype. Then for an interval $\Delta t$ of time, an
individual of genotype $x$ will be killed with probability
$1-\exp(-V(x)\Delta t)$ if $V(x) > 0$; if instead $V(x) < 0$, a clonal
birth will be produced with probability $\exp(-V(x)\Delta t) - 1$.
Following standard practice, if one wants
to keep the population size fixed on average near some
target value $M$, one simply shifts additively $V(x)$ so that the
expected population size is precisely $M$.
\end{itemize}

In the $M \to \infty$ limit, the details associated with keeping the
population at its target value no longer matter and the overall
process can be formulated as a rate equation. Up to the rescaling of
the population to keep its size fixed, the density of genotypes
$\rho(x)$ follows the deterministic equation
\begin{equation}
    \label{eq:schroedinger}
    \frac{\partial \rho(x,t)}{\partial t} = D
    \frac{{\partial}^2 \rho(x,t)} {\partial x^2} -V(x) \rho(x,t) .
\end{equation}
This is a linear evolution law that is the
continuum analog of the quasi-species dynamics.
At large times the \emph{shape} of
$\rho(x)$ converges to the eigenfunction of the linear operator on
the right hand side whose eigenvalue is largest. One can recognize
\eref{eq:schroedinger} as being an (imaginary time) Schrödinger
equation. The problem of the steady-state distribution is thus
mathematically a simple one that can be solved analytically for
particular choices of the function $V(x)$. Now we can address the
question of \emph{how} the large $M$ limit is reached.

\subsection{Harmonic well}
We first consider  the case where $V(x)= x^2$, which corresponds in
the Schrödinger equation framework to diffusion in a harmonic well,
a case with no barriers. In the $M\to \infty$ limit, the probability
distribution of the population in this landscape is
$P(x)=\exp(-x^2/2)/\sqrt{2 \pi}$.

\begin{figure}[htb]
\begin{center}
\begin{tabular}{cc}
\includegraphics[width=7cm]{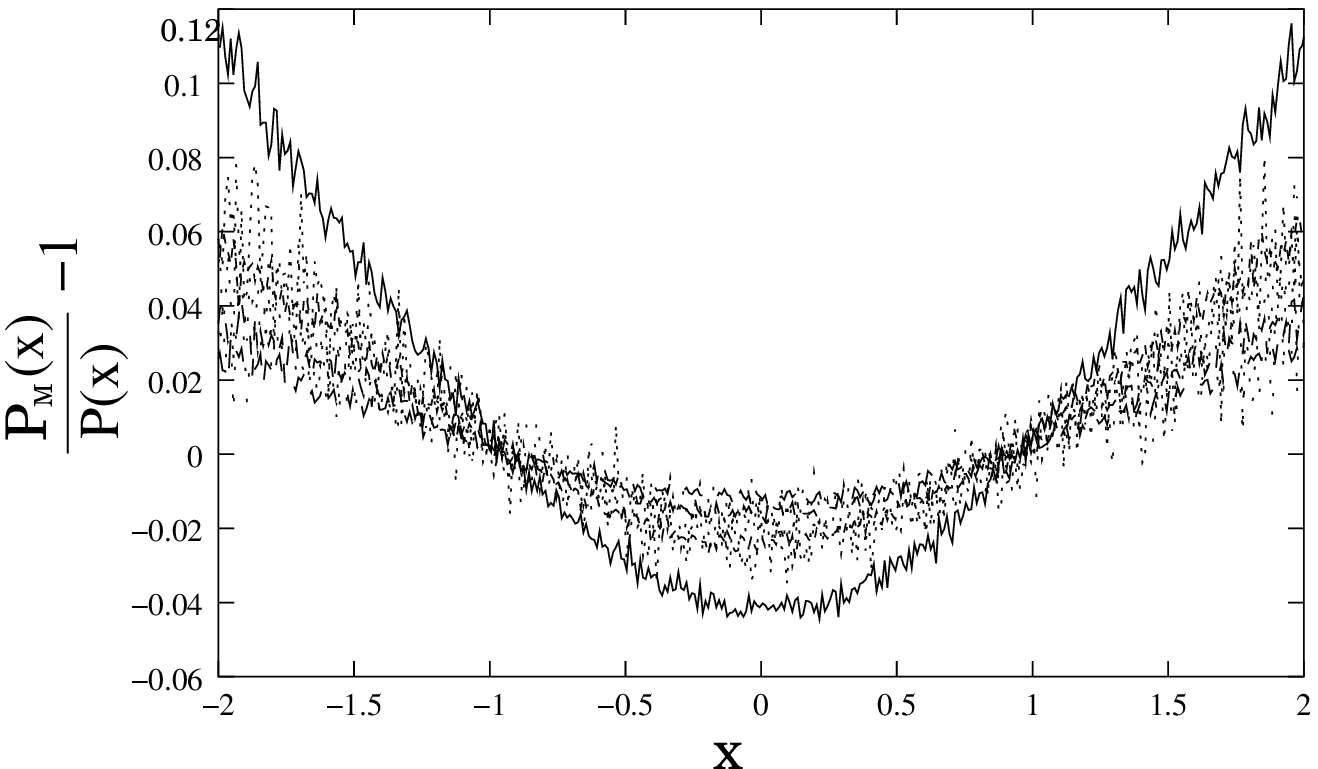} &
    \includegraphics[width=7cm]{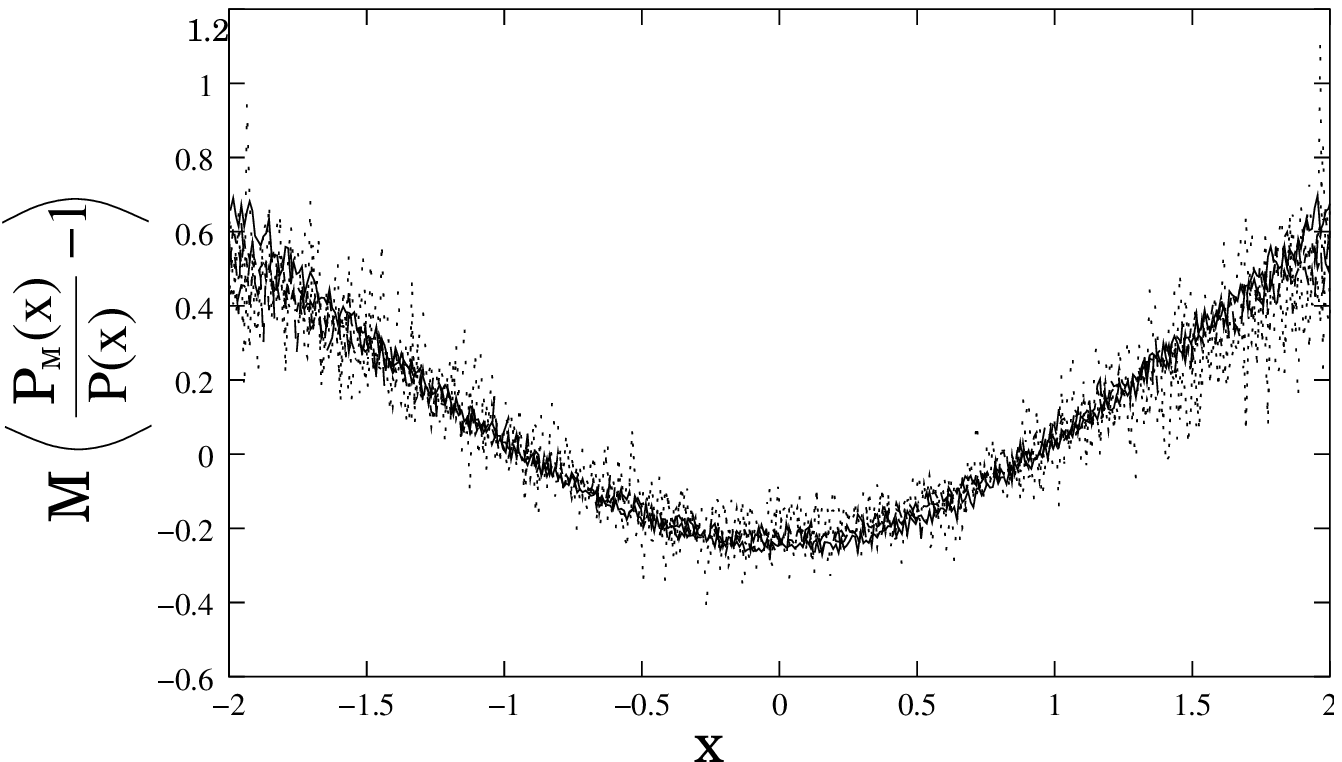} \\
\mbox{\bf (a)} & \mbox{\bf (b)}
\end{tabular}
\end{center}
\caption{(a) Relative deviation $P_M(x)/P(x)-1$ vs $x$ for
$M=6,8,10,12,15$ and 20. (b) Data collapse plot: $M(P_M(x)/P(x)-1)$ for
$M=6,8,10,12,15$ and 20.}
\label{fig:HarmonicWell}
\end{figure}

For a finite population of size $M$, we evaluate numerically the
steady state distribution $P_M(x)$. We see a clear convergence of
this distribution to its large-$M$ limit, with $1/M$ corrections:
\begin{equation}
P_M(x) = P(x)\left(1+\frac{K(x)}{M}\right) +
\Or\left(M^{-2}\right).
\end{equation}
The $1/M$ nature of the convergence clearly appears in
figure~\ref{fig:HarmonicWell}, where we see that the amplitude of
these corrections is small. This kind of convergence has been
justified before~\cite{CerfMartin95} in the context of population
algorithms for solving linear evolution operators. Furthermore, it
is possible to show that the correction function $K(x)$ goes to a
constant at large $x$. When we obtain such a data collapse, we know
$M$ is large enough for one to extract $P(x)$.

\subsection{Symmetric double well}
\begin{figure}
\begin{center}
\includegraphics[width=7cm]{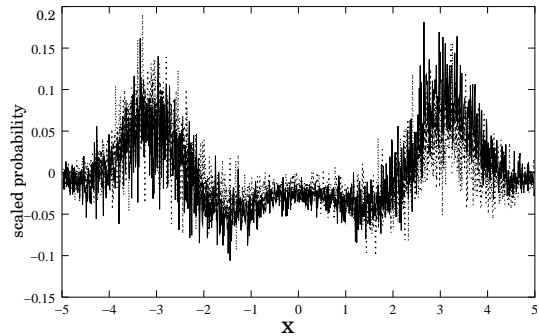}
\end{center}
\caption{Data collapse plot for $M_1 M_2 (P_{M_1}(x) - P_{M_2}(x))/(M_1
- M_2)$ vs $x$, for several population sizes $M_1$ and $M_2$, in
the case of a symmetric double well. We have taken $(M_1,M_2)=
(6,10),(10,20),(6,20),(20,50)$}
\label{fig:DoubleWell}
\end{figure}

We now consider the case of a landscape with two degenerate optima
separated by a barrier (passage of low fitness). For that we take
$V(x)$ to be an even polynomial of degree 4 in $x$: $V(x)=V(-x)$.
Since $P(x)$ is not analytically known, we examine instead the
quantity $M_1 M_2 (P_{M_1}(x) - P_{M_2}(x))/(M_1 - M_2)$ for
different population sizes $M_1$ and $M_2$. If the convergence goes
as $O(1/M)$, then, provided $M_1$ and $M_2$ are sufficiently large,
the data should collapse onto a limit function. This is indeed what
we find; the case $V(x) = x^2 + 0.1 x^4$ is used for illustration in
figure~\ref{fig:DoubleWell}.

\subsection{Case of an asymmetric double well}
To go from one fitness peak to another, one has to cross a barrier.
The previous double well has a symmetric steady state, and even with
a small population, this symmetry is realized. However when
the well is asymmetric, finite population effects will be quite
larger. Indeed, in the $M\to\infty$ limit, even a small non
symmetric part ($V(x)$ not even in $x$) will lead to a distribution
practically concentrated in one well. (This is called the ``flea and
elephant phenomenon'' well known in
quantum mechanics: when the barrier between the two wells is
high, even a tiny difference in $V$ between the two sides leads to a
big effect, just as when a little itching on an elephant's shoulder
leads it to put all of its weight onto one side.) When the
population is finite, this effect is not so evident, and the two
wells remain nearly symmetrically populated. One has to go to large
population sizes $M$ in order to come closer to the $M\to\infty$
limit.

\begin{figure}[htb]
\begin{center}
\includegraphics[width=8cm]{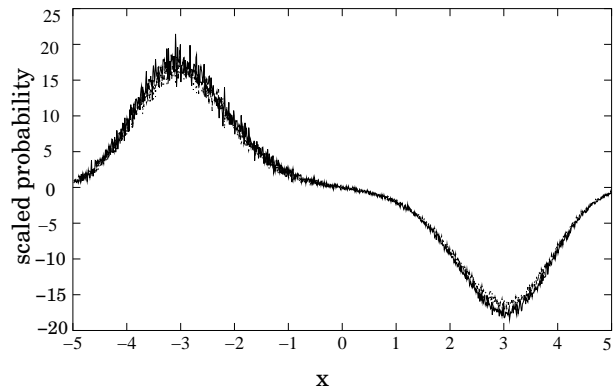}
\end{center}
\caption{Data collapse plot for
$M_1 M_2 (P_{M_1}(x) - P_{M_2}(x))/(M_1 - M_2)$ vs $x$, for
several population sizes $M_1$ and $M_2$, in
the case of an asymmetric well. $M=200,250,300,400,500$. }\label{asym:fig}
\end{figure}

To investigate this effect quantitatively, we consider the
asymmetric potential $V(x)=-x^2+b x^4+cx$ with $b=0.05$ and
$c=0.002$, corresponding to an asymmetry of roughly $0.1\%$. We
found the $1/M$ convergence law,
but had to go to $M \approx 100$ to observe
the data collapse. This is shown in figure~\ref{asym:fig}; note the
large scale of the $y$ axis compared to the symmetric double well
case: the $1/M$ corrections are much larger here. When the barrier
height is increased, one needs even larger $M$ values to see the
$M=\infty$ limit: the proper balance of population on each side of
the barrier sets in very slowly in $M$. In landscapes with more than
one dimension, there can also be entropic barriers, that is passages
that are narrow but not of particularly low fitness; such cases are
relevant for general neutral networks.

\section{Summary and conclusions}
\label{sec:Conclusions}

In general, an evolving population undergoes mutation, selection and
random drift. In this work we quantified the effect of these
processes to untangle the different effects, using neutral networks
and toy fitness landscapes for illustration. In the case of infinite
populations, there have been many studies. If $M \mu\to\infty$, ($M$
being the population size and $\mu$ the mutation rate), drift is
absent and one recovers the quasi-species
limit~\cite{Eigen71,EigenSchuster77,EigenMcCaskill89}, with results
that developed by van Nimwegen et al.~\cite{NimwegenCrutchfield99}
in the context of neutral networks. For finite $M \mu$ ($M \to
\infty$), drift effects are important; there, in the absence of
selection, Derrida and Peliti~\cite{DerridaPeliti91} derived a
number of important results. In this work we have considered the
case $M \mu$ finite with selection, for both $M$ finite and
infinite. We derived the $M \mu$ scaling even
in the presence of selection.
The (finite $M$) \emph{corrections} to
this $M \mu$ scaling are $O(1/M)$, be there selection or not.
When $M=\infty$, we find that the quasi-species limit is reached
via $1/M \mu$ corrections. In all cases,
$M \mu$ plays the role of an effective population size. These
laws are summarized in equations~(\ref{eq:CorrectionsToScaling})
and (\ref{eq:CorrectionsToInfinity}). We also
found that the amplitude of the correction terms showed a slow
increase with neutral network size. In practice, the $M \mu$ scaling
sets in at relatively small values of $M$. Furthermore, at fixed $M
\mu$ we showed that the genotypic diversity $g_M$ of the population
increases only logarithmically as a function of $M$.

Finally, we considered a dynamics with low drift in
section~\ref{sec:SelectionLowDrift}. Drift is effectively reduced by
a factor $\mu$ and thus genotypic diversity
always  grows \emph{linearly}
with population size even if $M \mu$ is fixed. One thereby avoids
the $M \mu$ scaling law, a useful property
if one wishes to evaluate the
large-$M$ limit at small $\mu$. Nevertheless, reaching this limit
can be seriously hindered by fitness or entropic barriers in the
fitness landscape as we saw in section~\ref{sec:Bottlenecks}.

\ack This work was supported by the EEC's FP6 IST Programme under
contract IST-001935, EVERGROW. We thank F. Hospital and A. Wagner
for their comments. LP is grateful to the LPTMS for its hospitality
when this work was started.

\section*{References}

\addcontentsline{toc}{chapter}{\protect\bibname}
\providecommand{\newblock}{}

\end{document}